\documentclass[conference]{IEEEtran}

\usepackage{graphicx,tabularx} 

\IEEEoverridecommandlockouts 

\begin{document}

\title{\ \\ \LARGE\bf Protein Sequencing with an Adaptive Genetic Algorithm from Tandem Mass Spectrometry 
\thanks{Jean-Charles Boisson, Laetitia Jourdan and El-Ghazali Talbi
are from the LIFL/INRIA Futurs, B\^at M3 
(email: {boisson,jourdan,talbi}@lifl.fr).}
\thanks{Christian Rolando is from the Plateforme de Prot\'eomique / Centre Commun de Spectrom\'etrie de masse,B\^at C4
(email: Christian.Rolando@univ-lille1.fr).}
\thanks{All the authors have the same address: (corresponding department, see above) 
Cit\'e Scientifique 59655 Villeneuve d'Ascq Cedex, FRANCE.}}

\author{Jean-Charles Boisson, Laetitia Jourdan, El-Ghazali Talbi and Christian Rolando}

\maketitle

\begin{abstract}
In Proteomics, only the \emph{de novo \textbf{peptide} sequencing} approach allows a partial amino acid sequence of a peptide 
to be found from a MS/MS spectrum. In this article a preliminary work is presented to discover a complete protein sequence 
from spectral data (MS and MS/MS spectra). For the moment, our approach only uses MS spectra. A Genetic Algorithm (GA) has 
been designed with a new evaluation function which works directly with a complete MS spectrum as input and not with a mass list like the 
other methods using this kind of data. Thus the mono isotopic peak extraction step which needs a human intervention is deleted.
The goal of this approach is to discover the sequence of unknown proteins and to allow a better understanding of the differences 
between experimental proteins and proteins from databases.     
\end{abstract}

\section{Introduction}

\begin{figure}[h]
\centerline{\includegraphics[width=3.35in]{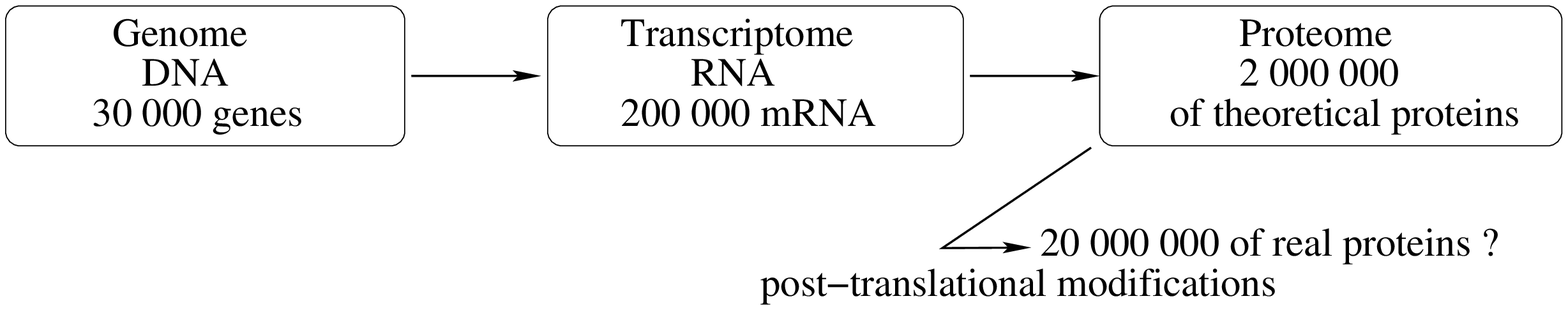}}
\caption{Global scheme: from genome to proteome (human case).}
\label{fig:proteom}
\end{figure}

Proteomics is a recent research domain which has emerged thanks to the mass spectrometry 10-15 years ago. It can be defined as 
the global analysis of proteins. The word proteome defines the protein set of an organism. Figure~\ref{fig:proteom} 
represents a global scheme starting from the genes to the proteins. 

The Proteomics main goal is the experimental protein identification. Several general techniques exist and a lot of identification
tools can be used to make experimental protein identification (the well known Mascot~\cite{APP1999} for example).
MS spectra is the most common data used to make a first step of identification. It is a mass/intensity spectrum where
each peak generally corresponds to a peptide of the experimental protein. A peptide is a subset of the original protein
obtained by a digestion mechanism. In this digestion a protein is cut at specific cleavage points by an enzyme. From a MS spectrum,
a mono isotopic mass list is extracted and used for the identification process. In order to make this identification, the 
Peptide Mass Fingerprinting (PMF) techniques take proteins from databases and theoretically digest them in order to suit with 
the experimental data. These methods compare the experimental mass list with theoretical mass lists and allow to find the protein 
which has the best score~\cite{AGMG1999,AMB2005}. The accuracy of these methods can be increased thanks to tandem mass
spectrometry. The tandem mass spectrometry corresponds to using MS/MS spectra in addition of MS spectra. 
A MS/MS spectrum is also a mass/intensity spectrum but each peak generally corresponds to one ion type. In 
fact, a MS/MS (or $MS^2$) is the result of the fragmentation of one peptide. So for each peptide, a MS/MS spectrum is generated.
These spectra give more information to identify close proteins than MS spectra~\cite{ABE2001,AMMC2004}.

An another way to use MS/MS spectra is to make identification by \emph{de novo sequencing}. Theoretically, if the ions
resulting of a peptide fragmentation can be all kept in the right order, the peptide sequence can be found. However,
the MS/MS spectra are noisy and only small sequences can be deduced. So the \emph{de novo sequencing} methods manage
to find the right peptide sequence to help the protein identification. These methods start from random peptide
sequences or from sequences gained by another identification tool in order to find the right peptide sequence
thanks to MS/MS spectra~\cite{ADAC1999,AFP2005,AMHBJC,ASD2004}. The main problem of this approach is the huge research
space of potential peptide sequences. Some optimization methods have been used with good results~\cite{AHLR2004,AMHBJC}.
Furthermore, the complete identification process need to be assisted by a sequence alignment tool like Blast to be complete. 

A complete automatizing of all the \emph{de novo peptide sequencing} process is interesting. But if a 
complete automatic \emph{de novo sequencing} approach can be used on all the peptides of one protein, a protein can be sequenced. 
Inspired by this idea, we propose a complete approach for making protein sequencing.

\begin{figure}[h]
\centerline{\includegraphics[width=3.35in]{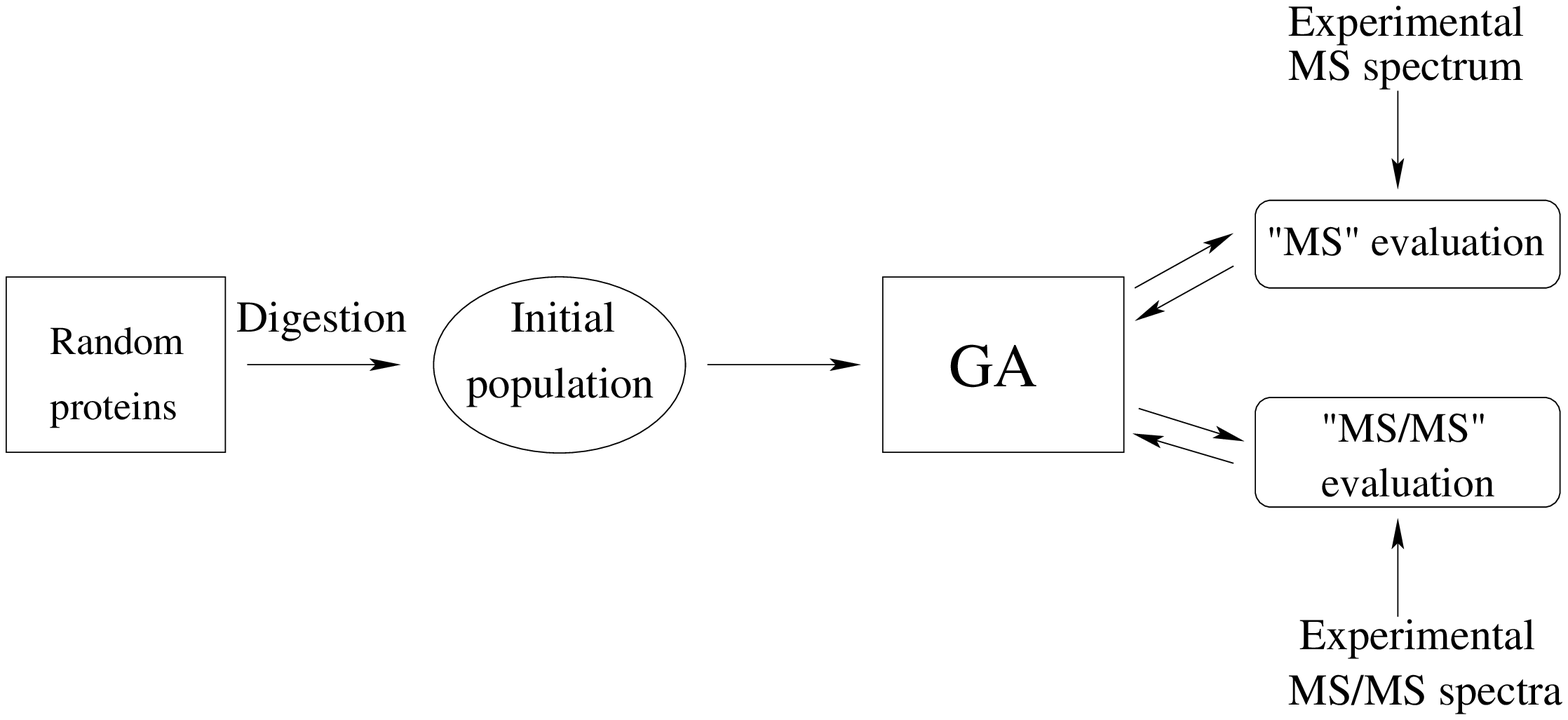}}
\caption{General approach scheme. ``MS''(``MS/MS'') evaluation: individual evaluation with a MS (MS/MS) spectrum. }
\label{fig:globalScheme}
\end{figure}

Figure~\ref{fig:globalScheme} illustrates this approach. From a MS spectrum (peptide level) and MS/MS spectra (ion level), 
the closest \textbf{protein} sequence may be generated. The originality of this work is that whereas all the other approaches 
use a list of peptide masses manually extracted thanks to a proprietary software from the spectrometer seller, we directly 
use a MS spectrum issued of the spectrometer. Furthermore discovering protein sequences is the only way to identify proteins 
unknown from databases. An other interest of protein sequencing is the possibility to detect sequence variations between the 
experimental protein and its representation in the databases. In this article, the first step of our approach that allows to 
find the experimental peptide chemical formula is presented. In
section~\ref{ga} each part of the chosen optimization method, 
a Genetic Algorithm (GA), is exposed. In section~\ref{results}, the statistics concerning the GA behavior and the first 
results of our approach is presented. Finally, section~\ref{conclusions} deals with the conclusions and perspectives 
about this work.   

\section{A specific genetic algorithm}
\label{ga}

\begin{figure}[h]
\centerline{\includegraphics[width=3.35in]{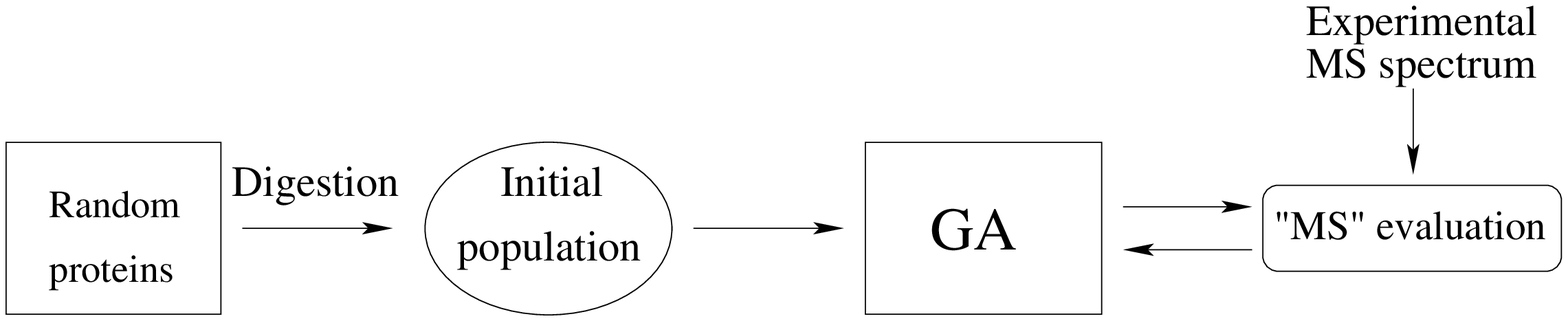}}
\caption{Actual approach scheme. ``MS'' evaluation: individual evaluation with a MS spectrum. }
\label{fig:actualScheme}
\end{figure}

In this section the global scheme of our approach is presented with an explanation of the digestion process. Then
each GA part is carefully described. Figure~\ref{fig:actualScheme} represents the actual version of our approach
in which only the MS evaluation is proposed.

\subsection{The approach}

In this part, the global approach and the theoretical digestion process are presented.\\

\subsubsection{Description}

a protein can be described as a sorted set of peptides. On the one hand a MS spectrum can help to globally identify the peptides 
which composed the experimental protein without any information on their sequences or their order. On the other hand a MS/MS 
spectrum which corresponds to a peptide can give information about the peptide sequence. So from MS and MS/MS spectra, the peptide
sequences may be available but the peptides may not be necessary in the right order. For the moment, a genetic algorithm has been 
designed with an new evaluation function working on a MS spectrum in order to find the chemical formulae of the peptides that compounds 
the experimental protein. Our evaluation function directly compares an experimental spectrum with a simulated spectrum generated 
from an amino acid sequence. This evaluation function may allow to find the right chemical formula (and so the right mass) 
of the peptides. In order to generate a simulated spectrum which can be compared with a MS spectrum, the analyzed protein has
to be in a peptide list format. To do that, a theoretical digestion has to be executed. The next paragraph describes
the digestion algorithm that we have designed.\\

\subsubsection{The Digestion Process}

in order to be analyzed, experimental proteins are cut by an enzyme before being put in the mass spectrometer: it is
the digestion step. There are several kind of enzyme and each of them cuts proteins on specific cleavage point. In fact, 
each enzyme respects its own cleavage grammar. For example, the trypsin enzyme cuts proteins after the amino acids lysine (K) and 
arginine (R) if they are not followed by a proline (P). However, in the real digestion process, enzyme can miss cleavage points and 
so the result peptides can have ``miss cleavage''. Due to these miss cleavages, the number of potential peptides that can be generated
 by the digestion process is increased. The developed theoretical digestion algorithm is an linear and iterative 
algorithm with no limitations in the number of considered miss cleavages. 

\subsection{the GA}

\begin{figure}[h]
\centerline{\includegraphics[width=3.35in]{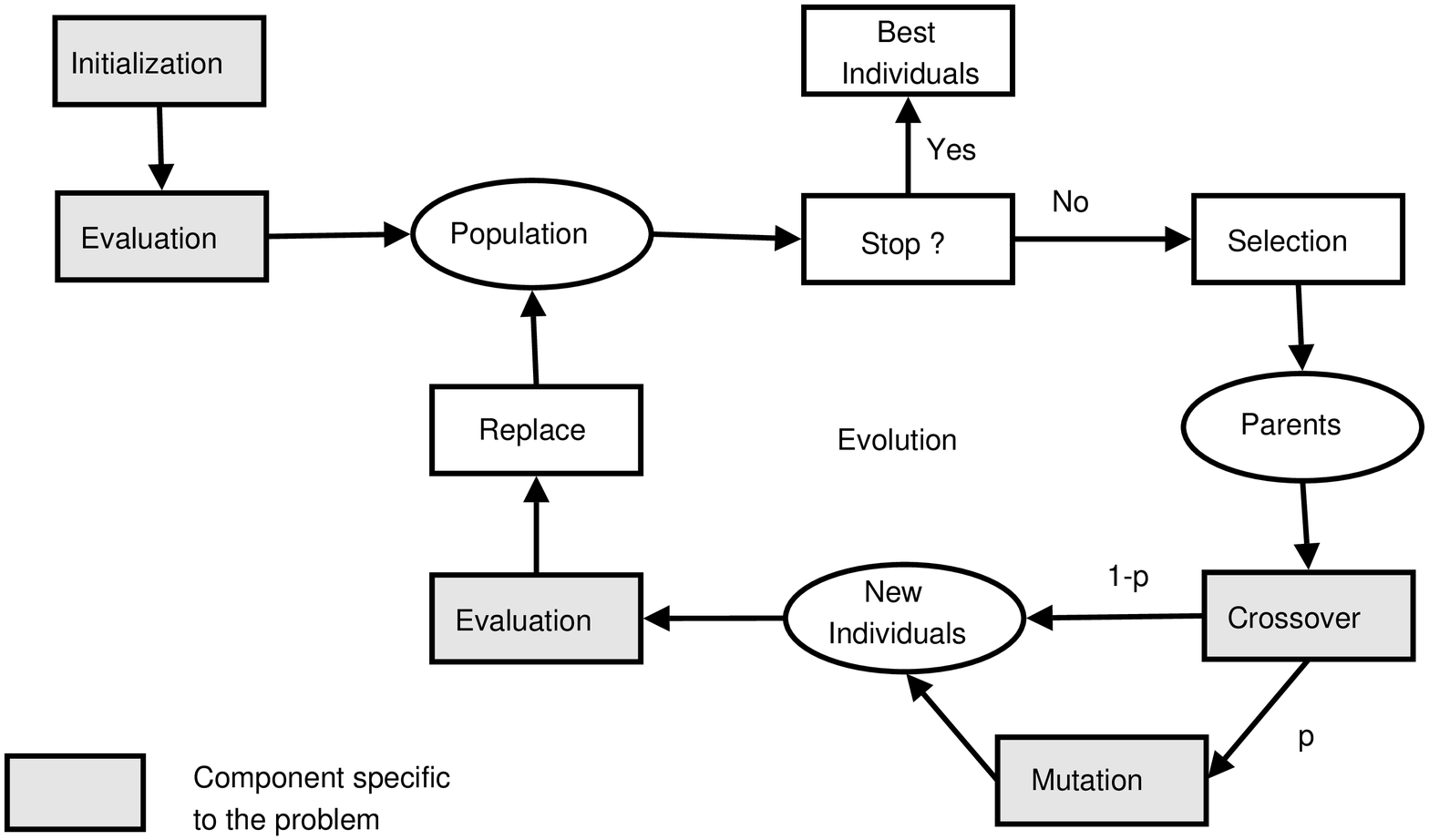}}
\caption{The general flowchart of a genetic algorithm. $p$ is the probability of mutation.}
\label{fig:aggen}
\end{figure}

In our approach, we want to sequence proteins. The search space linked to this goal can be described as follow: according to the 
size of a protein in amino acids ($n$) and the number of existing amino acids ($20$), there are $20^n$ potential proteins that 
can be generated. Nevertheless $n$ is unknown. Generally it is in $[100,10000]$ amino acids. However, bounds can be computed in order 
to reduce this range according to the experimental protein mass (see ``population initialization'' part). If we add static 
and variable post translational modifications, the number of potential proteins (already huge) explodes. So we need an optimization 
method that can work on very huge search space. That is why a genetic algorithm has been chosen. The initial protein population evolves 
according to specific crossover and mutation operators~\cite{hollan:ga}. Figure~\ref{fig:aggen} shows the global scheme of a GA.

Each part of the GA has been designed for our problem.\\ 

\subsubsection{Encoding}

For an individual, there are three possible manner to encode it. An individual could be:\\
\begin{itemize}
\item{An amino acid chain: it is the simplest representation. At each evaluation, the individual need to be digested 
(MS evaluation) and fragmented (MS/MS evaluation).\\}
\item{A peptide list: an individual is digested one time at its initialization, only miss cleavage computation need to be updated
when an operator modifies an individual. From the peptide list, it is easy to return to the original amino acid chain. However, 
to be evaluated with a MS/MS evaluation function, the peptide list needs to be fragmented.\\}
\item{A ions list: a MS/MS evaluation is direct but returning to peptide level or protein level is very difficult.\\}
\end{itemize}

Finally, the second representation has been chosen because it is easy to return to protein level and MS evaluation is direct.
An individual is a list of peptides (with the number of miss cleavage), each peptide is an amino acid chain and each amino acid can 
have post-translational modifications.\\ 

\subsubsection{Population initialization}

population is randomly initialized with a variable size (in amino acids). This size is contained between two bounds which are calculated
from the estimated experimental protein mass. From this mass, we compute a minimum protein mass and a maximum protein mass (experimental
mass $\pm$10 percents). The upper (lower) bound is calculated thanks to the maximum (minimum) protein mass and the amino acid 
that has the smallest (biggest) mass. So a maximum range for the protein size (in amino acid) is obtained. Nevertheless the generated 
protein mass has to be checked in order to be validated.\\

\begin{figure}[h]
\centerline{\includegraphics[width=3.35in]{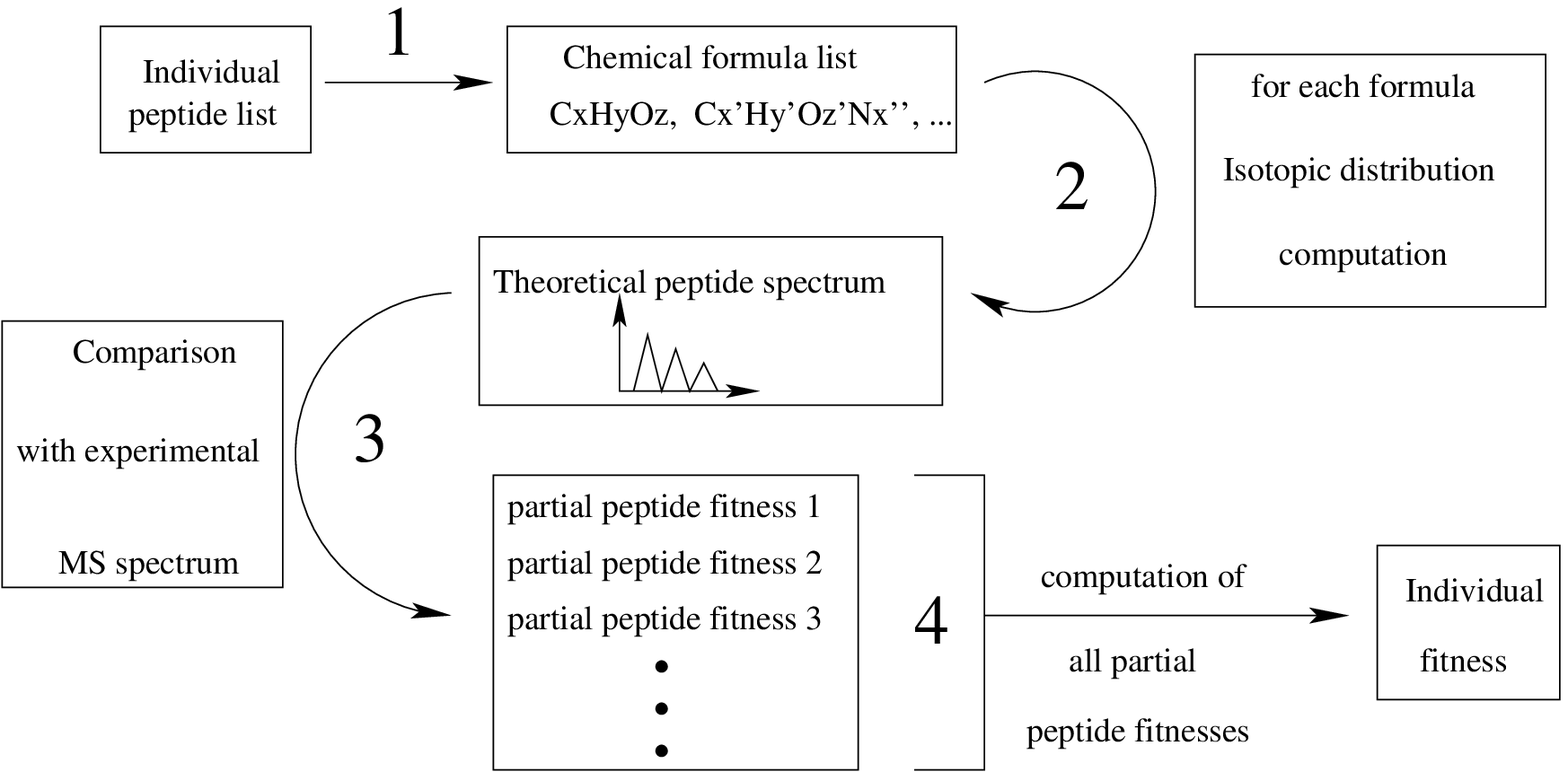}}
\caption{Evaluation of an individual.}
\label{fig:evaluation}
\end{figure}

\subsubsection{Fitness function}

the evaluation function compares an individual, transformed into a theoretical MS spectrum, with an 
experimental MS spectrum. A major interest of this function is to compare a MS spectrum with a simulated one (peptide by peptide).
The evaluation function does not need a mono isotopic mass list extracted from the experimental MS spectrum.
In order to generate a simulated spectra, we design a spectrum generator based on a algorithm developed by A.L. Rockwood~\cite{ARVO1995}
to compute isotopic distribution. For detail our fitness function, we use the following notations: $n$ is the protein size in
amino acids, $m$ is the protein number of peptides, $n_a$ is the number of elements in a chemical formula, $n_{xq}$ is the 
quantity of element X in a chemical formula and $N$ is the array size that contains a spectrum. $N$ is a very important parameter 
as its value sets the number of points that describes the spectrum. The higher N is, the more accurate is the spectrum. 
The used Fast Fourier Transform (FFT) algorithm is in $Nlog_2N$.
Figure~\ref{fig:evaluation} details the 4 steps to evaluate an individual.

We begin from an individual i.e. from a peptide list:\\
\begin{enumerate}

\item{The peptide list is transformed into a chemical formula list. This step is linear in the protein size: $O(n)$.\\}

\item{Each chemical formula allows to generate a part of the complete simulated spectrum. To do that, the
isotopic distribution of each formula is computed. For an element X, its initial isotopic distribution is computed ($X_1$ in
$O(N)$). Then FFT allows to change to the Fourier space ($O(Nlog_2N)$). In order to find the isotopic distribution for
$X_q$, $n_{xq}$ multiplications are needed ($O(n_{xq}*N)$). The isotopic distributions of each element has to be added 
together ($O(n_a*N)$). Finally, FFT allows to return to the Euclidian space ($O(Nlog_2N)$). So this step is in:
$$O(N+2Nlog_2N+n_a*n_{xq}*N+n_a*N)$$
$$\Leftrightarrow O(N*(1+2log_2N+n_a(n_{xq}+1)))$$
$$\approx O(Nlog_2N),~1 \ll n_a*n_{xq} \ll N$$
By default, $N$ has a value of $65536(2^{16})$.\\}

\item{Each part of the simulated spectrum (so each peptide spectrum) is compared with the experimental MS spectrum. A
partial score associated to a peptide is calculated. The peptides are classified according to their score:

\begin{itemize}

\item{Positive score: good correlation. The peptide appears in the two spectra and the isotopic distribution is
very similar (Figure~\ref{fig:peptide_correlation}, case A).\\}

\item{Negative score: bad correlation. There is maybe a peptide in the experimental spectrum but it is not 
similar to the theoretical peptide (Figure~\ref{fig:peptide_correlation}, case B).\\}

\item{The lowest scoring bound: no correlation. There is nothing in the experimental spectrum (Figure~\ref{fig:peptide_correlation}, 
case C). The lowest scoring bound is dynamically computed according the evaluation function configuration.\\}
\end{itemize}

\begin{figure}[h]
\centerline{\includegraphics[width=3.35in]{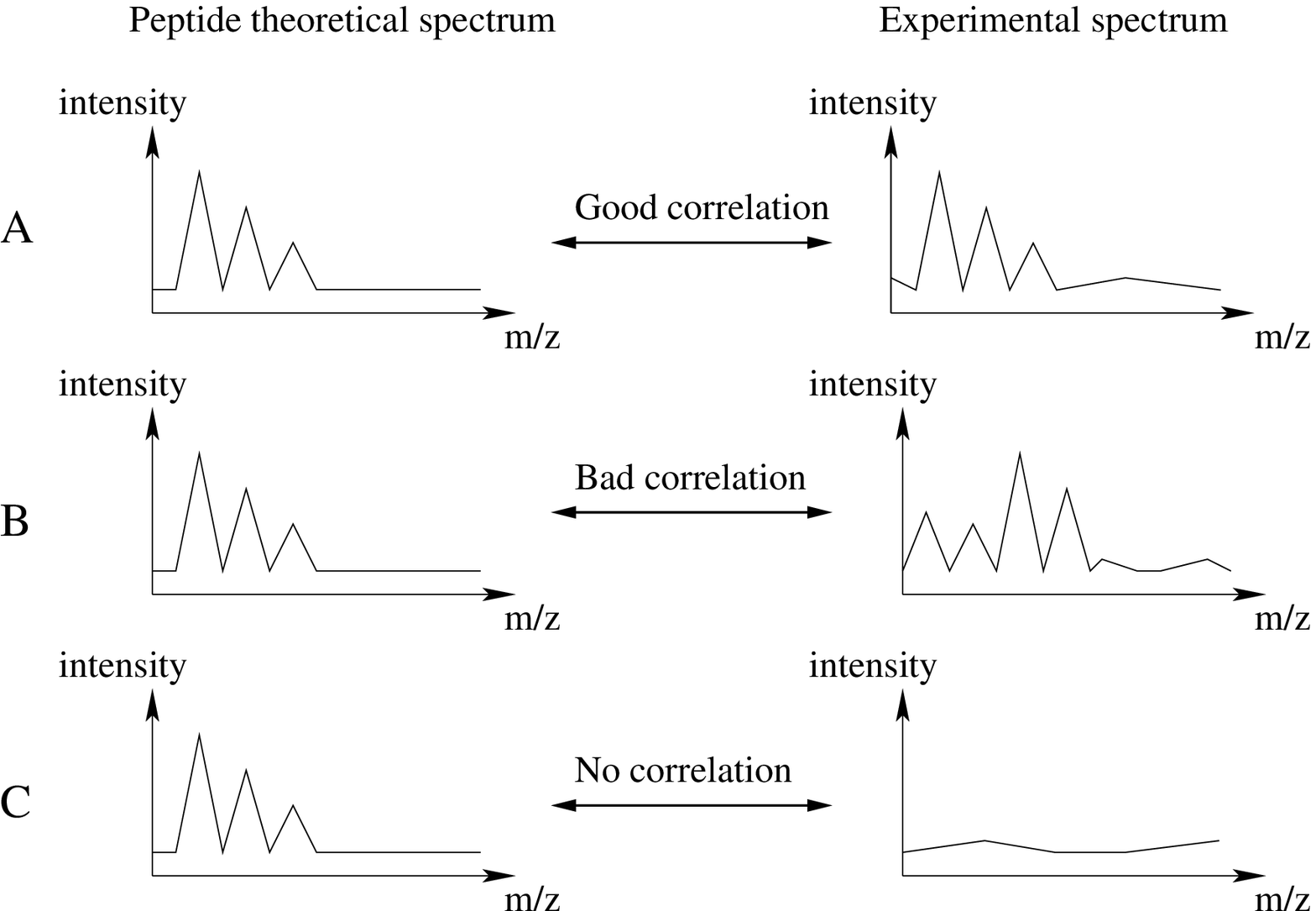}}
\caption{Correlation between a theoretical peptide and the experimental spectrum. case A: good correlation, case B: bad correlation and
case C: no correlation. m/z = mass / charge.}
\label{fig:peptide_correlation}
\end{figure}

This step is linear in the spectrum point number ($O(N)$).\\}

\item{The complete theoretical spectrum is finished. Using the peptide partial scores and the global similarity
between the two spectra give the individual fitness. This final step is also linear in the spectrum point number($O(N)$).\\}

\end{enumerate}

The original global complexity of our evaluation function is:
$$O(n+m*(Nlog_2N+N)+N),~m \ll n \ll N$$

Step 3 and 4 are computed with step 2, so the complexity become:
$$O(n+m*Nlog_2N),~m \ll n \ll N$$

The complexity is not very high but can be very time consuming due to the $N$ value. In order to
increase the speed of an individual evaluation, the isotopic distributions of the most common element are
computed for a range of atomic quantity. So the multiplication of the step 2 are no longer needed. The speed is increased but 
the allocated memory also. 

An individual evaluation is a resource and time consuming process. For example on a Pentium4 1.9Ghz,
a protein of 500 amino acids needs in average one second and 300 Mo of memory (constant value no linked to the protein size, see above) 
to be evaluated in the default evaluation function configuration. In this configuration, the theoretical spectrum is represented
by 65536 points ($2^{16}$). It can simulate peptides with a mass contained in the interval $[0,4096]$ (in Da). The accuracy
of the spectrometer is considered to be $10^{-6}$. The atomic isotopic distributions already calculated are only based
on the C atom quantity used (here $1000$), the other atom quantities are deduced from it ($X_q$ means X atom quantity):
$$H_q=4*C_q~~~~~N_q=C_q/2$$
$$O_q=C_q/4~~~~~S_q=C_q/8$$

These coefficients have been proposed by the proteomics platform chemists. Thus the isotopic distributions
for $C$ from $C_1$ to $C_q$, for $H$ from $H_1$ to $4*C_q$, \dots are computed in the evaluation function initialization. 
That is why there are 300 Mo of memory reserved, the most part is due to the different isotopic distributions which are 
already computed.   

This evaluation function has been validated by testing it as a simple protein identification tool by PMF. We
use the UNIPROT database in FASTA format that can be download at \emph{www.expasy.uniprot.org/database/download.shtml}.\\

\subsubsection{Operators}

There are two types of operator: the crossover and mutation operators. The crossover operator allows
from selected ``parents'' to generate ``children''. The mutation operators make small modifications
on the individuals to keep a genetic diversity in the population.

The chosen crossover is the well known 1-point crossover. Two individuals are selected (the parents), a cut point
is randomly placed at the same position in the two individuals and they exchange all the information positioned after this cut point.
Two new individuals (the children) are obtained, they have information from the two initial individuals but they are different.

Six mutation operators have been designed:
\begin{itemize}

\item{The peptide insertion: a randomly generated peptide is inserted in the peptide list that represents an individuals.
The size of the new peptides (in amino acid) corresponds to the average size of all the peptides that compounds the
individuals. This mutation may allow to reach new interesting peptides.\\}

\item{The peptide deletion: A randomly chosen peptide is deleted. This mutation may allow to increase the individual quality by
removing a peptide that penalizes the individual fitness.\\}

\item{The amino acid insertion: a random amino acid is inserted in a peptide of the individual peptide list. This new
amino acid may not generate a new miss cleavage. This mutation increases the peptide size.\\}

\item{The amino acid deletion: a randomly chosen amino acid is deleted of a peptide of an individuals. As the amino acid
insertion, this mutation modifies the peptide size. With this mutation, the peptide size is decreased.\\}

\item{The amino acid substitution: a randomly chosen amino acid is replaced by another one not randomly chosen. The
new amino acid is taken according a probability linked to the initial amino acid which is replaced. This probability
comes from a substitution matrix that gives for each amino acid, the probability to be replaced by another amino
acid. The default matrix used is the BLOSUM62 matrix~\cite{AHH1992} but others matrix can be specified. This mutation
may allow to modify the chemical formula of the peptide without changing its size (in amino acid).\\}

\item{The post-translational modification: a post-translational modification is added on a global peptide or on a amino acid 
according to the modification. The post-translational modifications are specific to proteins and are very important
in the protein activity. Some proteins are only activated thanks to post-translational modifications.\\}
\end{itemize}    

\begin{figure}[h]
\centerline{\includegraphics[width=3.35in]{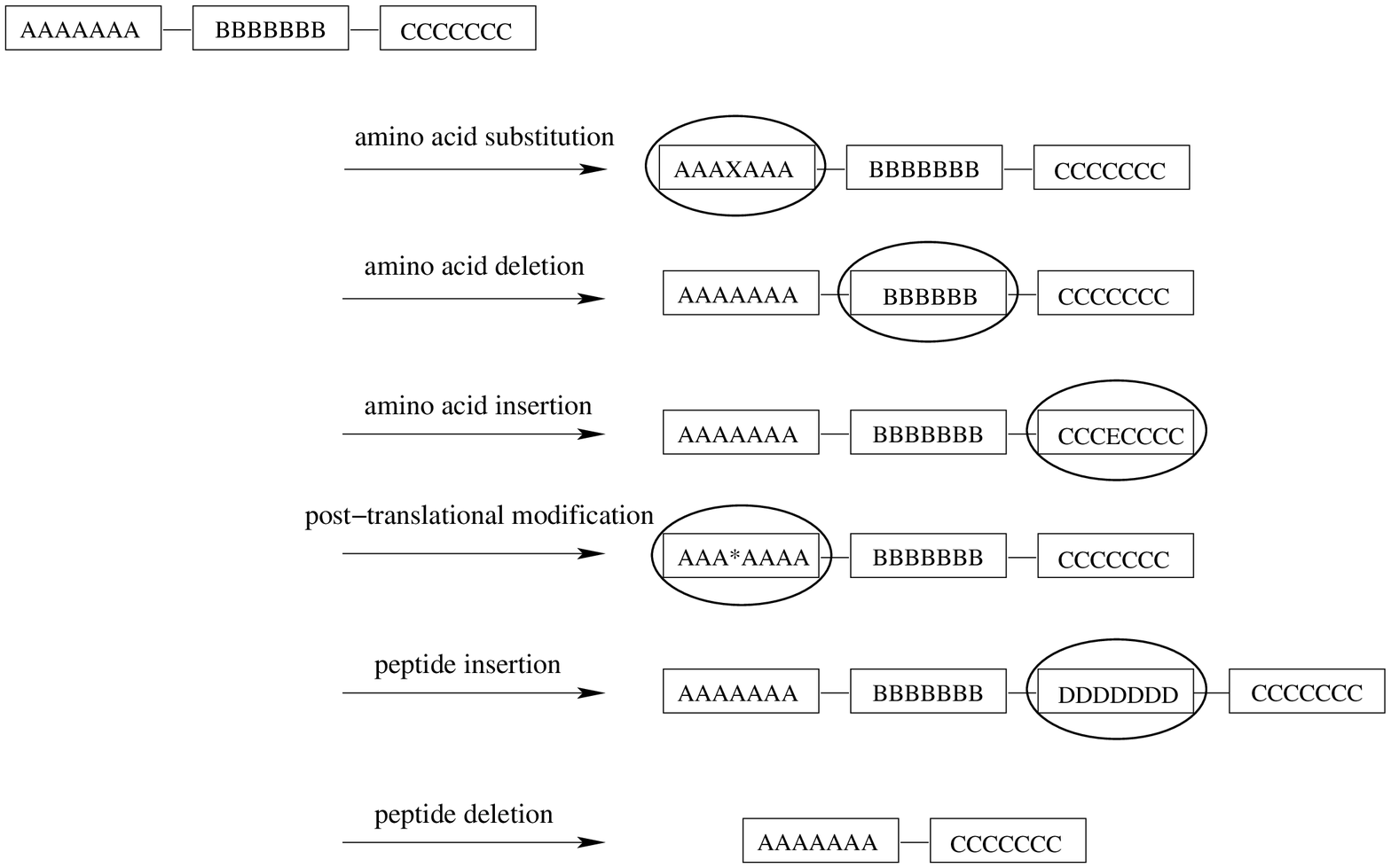}}
\caption{The mutation operator: the six types of mutation.}
\label{fig:mutOperators}
\end{figure}

All these mutations, summarized in figure~\ref{fig:mutOperators}, can be classified in two groups: the ``tiny'' 
mutations (amino acid insertion/deletion/substitution and post-translational modification) and ``small'' mutations 
(peptide insertion/deletion). The ``small'' mutations have a bigger impact on the individuals fitness than the ``tiny'' mutations.

As we have six types of mutation, it is difficult to set the probability of each of them. 
To overcome this problem, we implement an adaptive strategy for calculating the rate of each
mutation operator. Many authors have worked on setting automatically probabilities of applying
operator~\cite{davis:adapt,julsrtom:adapt,herrera96adaptation}. In~\cite{mut:hong}, authors proposed 
to compute the new rate of mutation by calculating the progress of the $j^{th}$ application of mutation
operator $M_i$, for an individual $ind$ mutated into an individual $mut$ as follows:
$$ progress_j(M_i)=Max(fit(ind), fit(mut))-fit(ind)$$

With this mechanism, we can evaluate the evolution of the impact of each mutation operator during
the GA execution.

Each part of the GA have been detailed. The next section presents the first results of our approach.

\section{Results}
\label{results}

In this section, the statistics concerning the GA behavior are presented according to different
configurations of the parameters. Then a first biological validation of our approach is proposed.

For all the experiments, we have used two types of data:\\
\begin{enumerate}
\item{Experimental MS spectra: these spectra have been given by the proteomics platform collaborator. They have been produced
by a MALDI-TOF mass spectrometer. They correspond to real data from current experiments.\\}
\item{Simulated MS spectra: these spectra are theoretical spectra we have generated from protein sequences in FASTA format.
These type of data are useful to make tests without noise and protein mix. They can be considered as easy instances for
our approach. Furthermore, we can generate a lot of data because only protein sequences are needed.\\} 
\end{enumerate}

Table~\ref{tab:data} summarizes the proteins used for our first tests.

\begin{table}[h]
\centering
\caption{Main data used for our experiments. Apo-AI: Apolipoprotein-AI, Cyt-C: Cytochrome-C. Exp: experimental, Sim: simulated.
The protein length is given in amino acids (aa) and the protein weight in dalton (Da).}
\label{tab:data}
\begin{tabularx}{3.35in}{|X|X|X|X|X|}
\hline
\centering{\textbf{Name}} & \centering{\textbf{specie}} & \centering{\textbf{Type}} &
\centering{\textbf{Size(aa)}} & \textbf{Weight(Da)}\\
\hline
\centering{Apo-AI} & \centering{Human} & \centering{Exp} & \centering{$\emptyset$} & $\simeq~36000$ \\ 
\hline
\centering{Apo-AI} & \centering{Human} & \centering{Sim} & \centering{317} &  36112.71 \\
\hline
\centering{Cyt-C} & \centering{Bovine} & \centering{Exp} & \centering{$\emptyset$} & $\simeq~11500$ \\ 
\hline
\centering{Cyt-C} & \centering{Bovine} & \centering{Sim} & \centering{104} &  11565.02 \\
\hline
\end{tabularx}
\end{table}

The experimental spectra have an estimated size deduced from their MS spectrum but we can not give an estimated length for
the experimental amino acid sequence as it is unknown. The simulated spectra have been generated from the
corresponding sequence in the UNIPROT database in FASTA format that can be download at 
\emph{www.expasy.uniprot.org/database/download.shtml}. So the simulated data sequence length and their weight are easily computed.
Take the same protein under the experimental and the simulated format may allow to understand the difficulties linked to 
experimental data (spectrometer calibration, noise, \dots).

\subsection{GA behavior}

\begin{figure}[h]
\centerline{\includegraphics[width=3.35in]{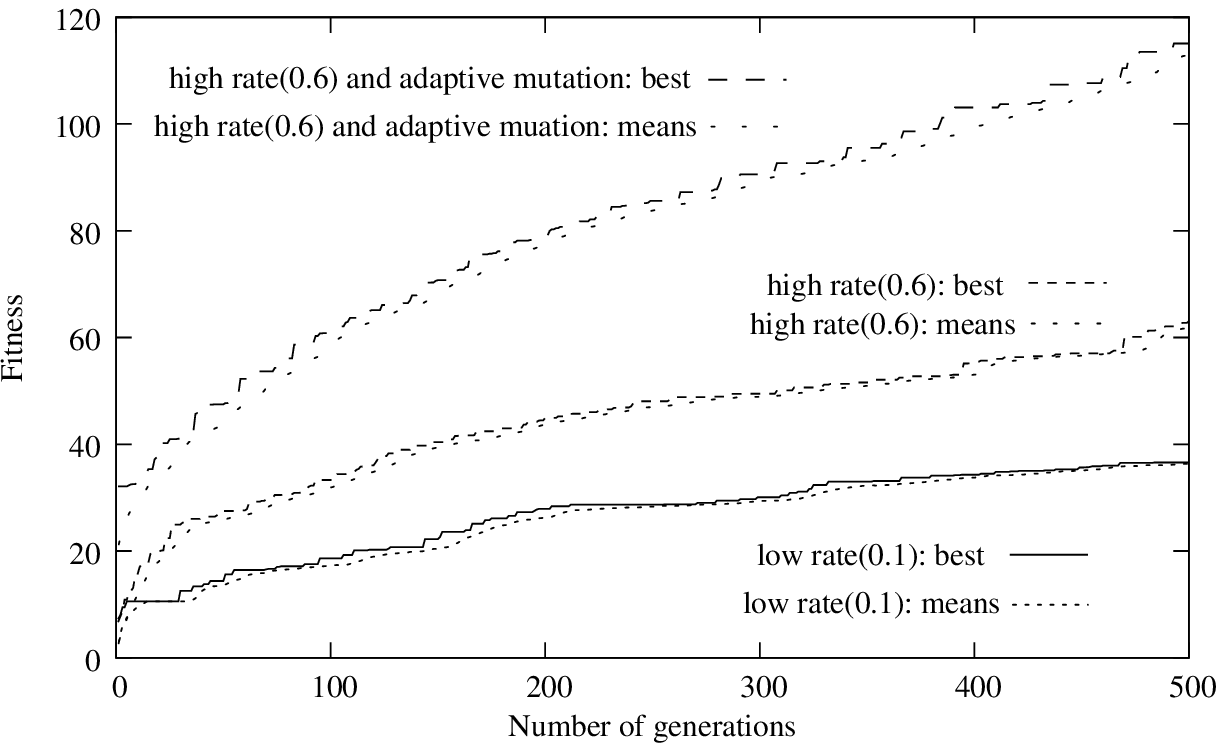}}
\caption{Evolution of GA convergence according to the mutation rate value and the adaptive mutation activation state.}
\label{fig:mutationRateAdaptEvolution}
\end{figure}

In order to validate our approach, the GA behavior have to be analyzed. According to the different version we
have developed for each part of our GA and all the different parameters, a lot of configuration can be realized. 
Each configuration test is time expensive due to the evaluation function. So we have developed a parallel version of the 
GA thanks to the ParadisEO platform~\cite{Cahon04a}. Due to the evaluation function cost, we have decided to parallelize 
the individual evaluation according to a master/slave scheme. The master initializes the population, the 
slaves evaluate it and at each generation, the master computes the crossover, the mutation and the population replacement steps
whereas the slaves compute the fitness of the new individuals. This version has been used on the French 
grid called Grid5000 (\emph{www.grid5000.org}). For each configuration test, we have made 15 runs to make first 
statistics on few generations (500). Concerning the crossover and mutation they have been selected as follow:

\begin{itemize}
  
\item{crossover rate: it has been set to 0.9 and experimentations have not shown the necessity to modify it.\\}

\item{mutation rate: it has been set to 0.1 and with this rate, the GA convergence was very slow. So we experiment several
rates of mutation in order to find the better one. Finally, we set this rate with a 0.6 value. In the following paragraphs, 
NAX GA will correspond to the GA with a mutation rate of 0.X without the adaptive mutation. AX will correspond to the 
GA with a mutation rate of 0.X with the adaptive mutation. Figure~\ref{fig:mutationRateAdaptEvolution} shows the convergence 
improvement with the Apolipoprotein-A1 example. Improve the mutation rate (NA6 curve) allows to obtain the same
quality of solutions than the NA1 GA in only 110 generations. At the end of the 500 generations, 
the individuals have a fitness value 2 times better than they have with the old mutation rate. We can also remark that the distance 
between the NA1 and NA6 curves is globally the same during the evolution. Thus, the gain is constant during the 500 generations. 
With this new mutation rate value, the GA behavior is better.\\}
\end{itemize} 

Furthermore, figure~\ref{fig:mutationRateAdaptEvolution} shows also the convergence improvement when the adaptive mutation is
activated. The final individual quality of the NA1 GA is obtained in:

\begin{itemize}
\item{only 110 generation for the NA6 GA.}
\item{only 20 generation for the A6 GA.}
\end{itemize}

We can remark that the distance between the A6 GA and the NA6 GA increases during the 500 generations. 
Concerning the final individual quality, comparing to the NA1 GA, the fitness is:

\begin{itemize}
\item{2 times better for the NA6 GA.}
\item{4 times better for the A6 GA. Furthermore, the increasing gain seems to continue in this case. }
\end{itemize}

For each configuration of your GA, the same statistics based on 10 runs will be exposed: the data used (experimental spectrum or 
simulated spectrum), the optimal fitness with the known protein corresponding to the MS spectrum (only in the 
case of simulated spectra), the best individual fitness, the fitness mean and the standard deviation. Table~\ref{tab:stats1} 
presents these statistics for the GA without adaptive mutation. 

\begin{table}[h]
\centering
\caption{GA statistics without adaptive mutation. AAI: Apo-AI Human, CC: Cyt-C Bovin.
 S/E: simulated/experimental spectrum.~$\sigma$: standard deviation.}
\label{tab:stats1}
\begin{tabularx}{3.35in}{|X||X|X|X|X|X|}
\hline
\centering{Data} & \centering{Max} & \centering{Best} & \centering{Mean} & \centering{Median} & ~~~~$\sigma$\\
\hline
\centering{AAI S} & \centering{78.47} & \centering{52.8574} & \centering{45.2751} & \centering{45.8461} & ~~6.4019\\
\hline
\centering{AAI E} & \centering{$\emptyset$} & \centering{47.7902} &  \centering{36.3106} & \centering{36.8451} & ~~7.3803\\
\hline
\centering{CC S} & \centering{19.4578} & \centering{15.5672} & \centering{11.3776} & \centering{11.324} & ~~2.56\\
\hline
\centering{CC E} & \centering{$\emptyset$} & \centering{44.5161} & \centering{27.1763} &  \centering{25.7919} & ~~7.7616\\
\hline
\end{tabularx}
\end{table}

We can remark that the protein size have an impact of the individual fitness because the fitness obtained 
for Apo-AI (experimental and simulated spectrum) is higher than the one gained with Cyt-C (experimental and simulated 
spectrum). Furthermore, the global statistics of our GA are better on simulated data than experimental data. That is due
 to different factor:

\begin{itemize}
\item{the spectrometer calibration: as we compare spectra, we estimate that the spectrometers are perfectly
calibrated. As the simulated spectra are ``perfect'' spectra, the GA behavior is better.\\}
\item{the spectrum noise: with experimental spectra, we have all the information but also noise can be present.\\}
\item{another proteins: when we gain a MS spectrum, there are not peptides from only one protein. There are always the 
possibility to have another protein peptides (from the enzyme used for the digestion for example).\\}
\end{itemize}

\begin{table}[h]
\centering
\caption{GA statistics of the experimental Apolipoprotrein-AI (AAI E) with adaptive mutation.}
\label{tab:stats2}
\begin{tabularx}{3.35in}{|X||X|X|X|X|X|}
\hline
\centering{Data} & \centering{Max} & \centering{Best} & \centering{Mean} & \centering{Median} & ~~~~$\sigma$\\
\hline
\centering{AAI S} & \centering{78.47} & \centering{\textbf{72.4086}} & \centering{62.8111} & \centering{68.0562} & ~~10.4471\\
\hline
\centering{AAI E} & \centering{$\emptyset$} & \centering{\textbf{120.752}} &  \centering{105.0315} & \centering{106.8990} & ~~12.6105\\
\hline
\centering{CC S} & \centering{19.4578} & \centering{\textbf{17.3078}} & \centering{14.0618} & \centering{14.6003} & ~~2.6220\\
\hline
\centering{CC E} & \centering{$\emptyset$} & \centering{\textbf{58.1147}} & \centering{45.1971} &  \centering{46.7804} & ~~8.7158\\
\hline
\end{tabularx}
\end{table}

Table~\ref{tab:stats2} shows the improvement of the experimental Apolipoprotein-AI when the adaptive mutation
is activated. In the four cases, the best individual fitness is increased.

\begin{figure}[h]
\centerline{\includegraphics[width=3.35in]{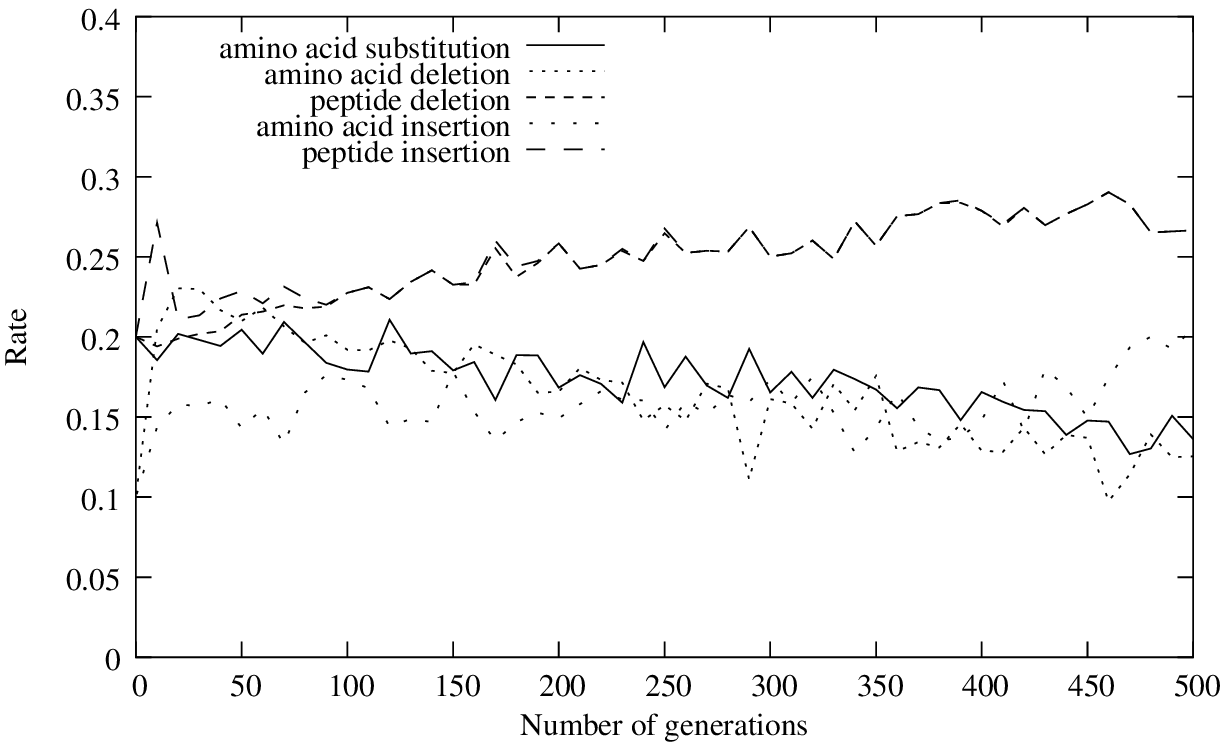}}
\caption{Evolution of the rate mutation according to each mutation (without post-translational mutation) with default
evaluation function configuration.}
\label{fig:probVar1}
\end{figure}

\begin{figure}[h]
\centerline{\includegraphics[width=3.35in]{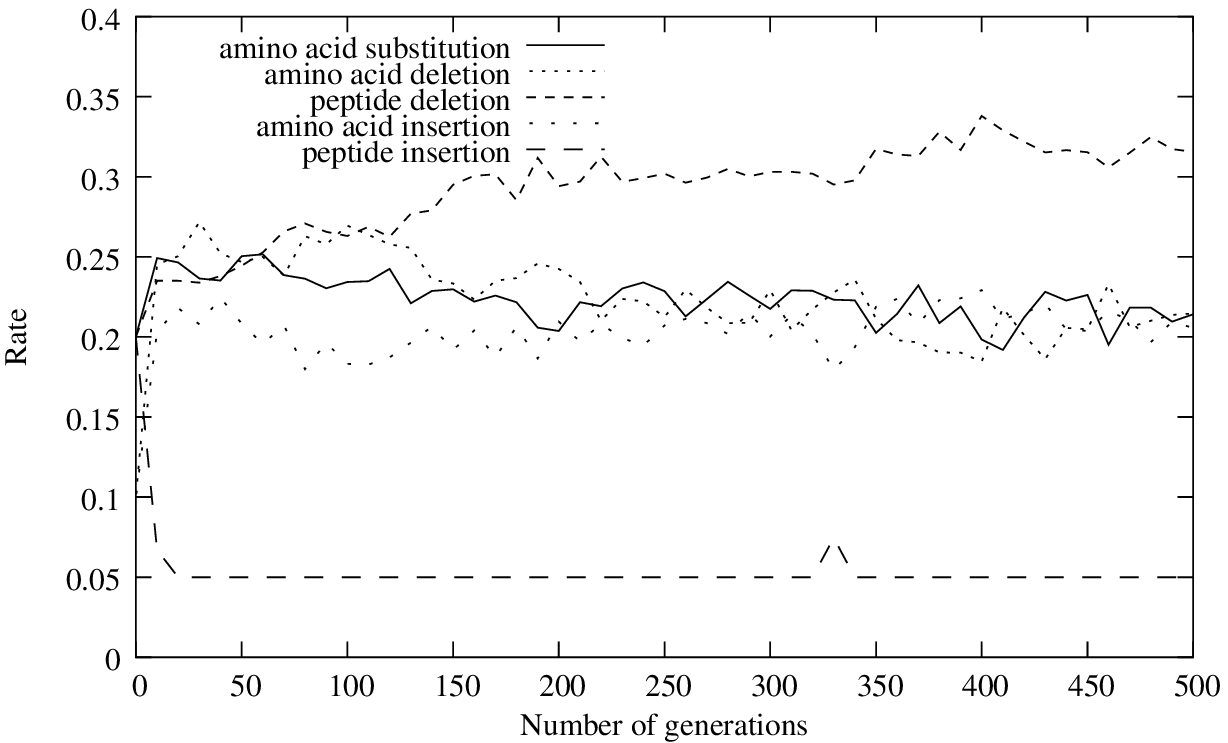}}
\caption{Evolution of the rate mutation according to each mutation (without post-translational mutation) with another
evaluation function configuration.}
\label{fig:probVar2}
\end{figure}

As the adaptive mutations are used, analyzing the operator mutation rate variation allows to understand how the GA evolves.
The GA evolution is directly linked to the used evaluation function. Figure~\ref{fig:probVar1} and~\ref{fig:probVar2}
show how the operator mutation rates move during the GA evolution for two configuration of the evaluation function.
The difference between these two configurations concerns only the coefficient used during the last step of an individual evaluation.

In figure~\ref{fig:probVar1}, the peptide insertion and the peptide deletion are the mutation operator the most used during the GA.
However, in figure~\ref{fig:probVar2} the peptide insertion is rapidly penalized whereas the peptide deletion is always the most
used mutation operator. Concerning the mutation rate for the operators working on the amino acids (amino acid substitution/deletion/
insertion), figure~\ref{fig:probVar1} shows that these operators have different evolution curves but globally their probability
of being used decrease during the GA evolution. On the contrary, in figure~\ref{fig:probVar2} these operators keep the same behavior
and their rate do not decrease nor increase. 

These results indicate that the configuration of the evaluation function greatly influences the GA behavior.

After the study of the GA behavior, the first results of the actual approach are proposed in the next part.

\subsection{Biological validation}

\begin{figure}[h]
\centerline{\includegraphics[width=3.35in]{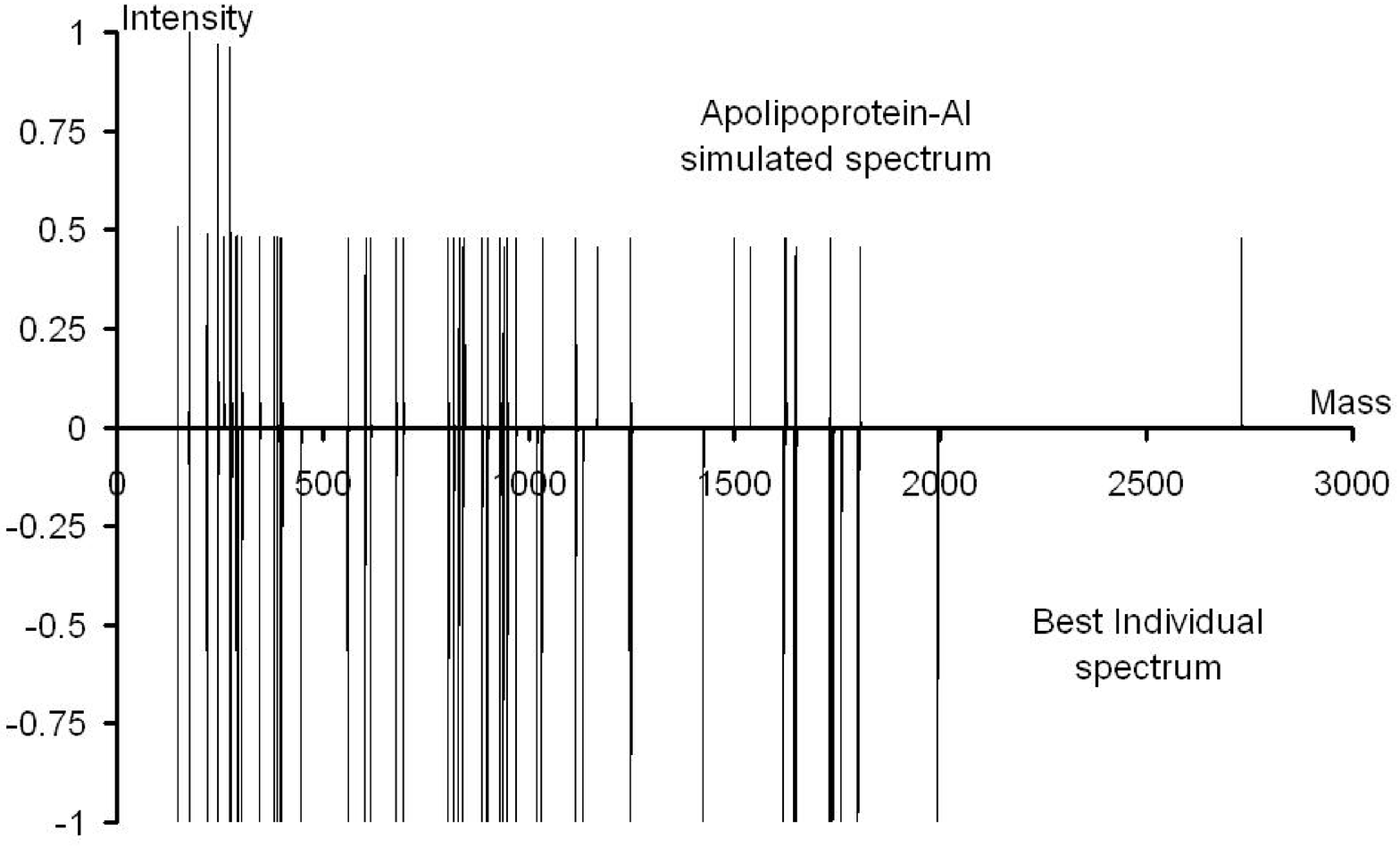}}
\caption{Apo-AI simulated spectrum vs best individual spectrum.}
\label{fig:spectra}
\end{figure}

As we have already explained, experimental spectra and simulated spectra have been used to test
the GA behavior. In the biological validation process, we used also these two types of data to evaluate 
the robustness of our result according to the spectrum quality. Our evaluation may allow to find
the right peptide chemical formula, so the best individual may have a spectrum very similar to the data used.
For example, figure~\ref{fig:spectra} shows the simulated spectrum of
one of the best individuals compared to the Apo-AI simulated one. For the moment only the place of the
peaks is analyzed, not the peak intensity because the spectrum
generation does not compute the peak intensities. A high intensity for
a simulated spectrum only indicates that several peptides have the
same mass. In figure~\ref{fig:spectra}, we remark that the same peaks are globally reached.

However, the peptide masses have also to be analyzed to be sure of the similarity. If the below example is
more precisely analyzed (show in table~\ref{tab:mass_matching}), the
individual peptides have the right mass for 11 of them or very close
mass for 20 of them. In some case, the right sequence is also found. However, having the right chemical
formula (so the right mass) does not seem that the same sequence has been found. Only
further work on MS/MS data can provide sequence information.  

\begin{table}[h]
\centering
\caption{Matching Apo-AI peptides (AAI pep) and best individual peptides. $\delta$ is the mass difference, $\delta$= ($|$Apo-AI 
peptide - best individual peptide$|/$Apo-AI peptide). There are also 11 exact sequence matches which are not shown here.}
\label{tab:mass_matching}
\begin{tabularx}{8.25cm}{|X|X|X|X|}
\hline
\centering{AAI pep} & \centering{$\delta$} & \centering{AAI pep} & ~~~~~$\delta$\\
\hline
\centering{278.153837} & \centering{9.03 $10^{-5}$} & \centering{839.339148} & ~~3.93 $10^{-5}$\\
\hline
\centering{347.229445} & \centering{2.9 $10^{-3}$} & \centering{886.474654} & ~~2.00 $10^{-5}$\\
\hline
\centering{381.213795} & \centering{3.2 $10^{-5}$} & \centering{899.441563} & ~~2.05 $10^{-5}$ \\
\hline
\centering{561.263264} & \centering{7.19 $10^{-5}$} & \centering{930.504892} & ~~1.07 $10^{-3}$\\
\hline
\centering{603.335367} & \centering{1.64 $10^{-3}$} & \centering{938.432714} & ~~9.94 $10^{-4}$\\
\hline
\centering{616.378235} & \centering{1.7 $10^{-3}$} & \centering{948.526690} & ~~1.18 $10^{-11}$\\
\hline
\centering{678.393885} & \centering{1.5 $10^{-3}$} & \centering{968.552905} & ~~6.25 $10^{-5}$\\
\hline
\centering{804.373937} & \centering{6.03 $10^{-5}$} & \centering{1114.585661} & ~~9.72 $10^{-4}$ \\
\hline
\centering{817.395676} & \centering{2.27 $10^{-5}$} & \centering{1247.576887} & ~~1.11 $10^{-5}$ \\
\hline
\centering{830.437206} & \centering{1.24 $10^{-3}$} & \centering{1647.801210} & ~~5.60 $10^{-6}$\\ 
\hline
\end{tabularx}
\end{table}

\vspace{5cm}

For the moment, our approach can not be compare for the moment to another tool for two main reasons:

\begin{itemize}

\item{our approach is not complete. The MS/MS evaluation level is not already implemented and it is the next
step to reach the protein sequence.\\}

\item{Only \emph{de novo peptide sequencing} approaches can be compared (with for example the Lutefisk tool). But 
our approach is designed for making \emph{de novo protein sequencing}.\\}

\end{itemize}

This approach may be used to give more information of possible protein sequence modifications.  

\section{Conclusions and perspectives}
\label{conclusions}

A first step for a fundamental approach to identify experimental
protein sequence has been proposed. We have designed a GA with a new
evaluation function that avoid a needed step in all the other methods
using MS spectra: the extraction of the mono isotopic list that needs human intervention
via a proprietary software linked to the used spectrometer. The first
tests have given interesting results. The individual result of the GA 
evolution has a MS spectrum closed to the experimental
one. Therefore, the right chemical formula are found. Furthermore, the size of each
peptide (in amino acids) is also correlated with the data. So the search space is reduced.

However, the peptides of the result individuals generally don't have the right sequence and they are not in the right order.
To overcome these problems, (1) MS/MS spectra can be used to find the right peptide sequence and (2) a MS spectrum of 
the experimental protein gained with another digestion enzym (for example pepsine) may allow to find the right peptide
order that gives the optimal fitness with the new MS spectrum. This approch has been validated by the
proteomics platform collaborator and is under implementation.  

Moreover, the study of the GA behavior has shown that the current crossover used
(the 1-point crossover) is not effective. So two main possibilities to
increase the GA behavior can be proposed:

\begin{itemize}
\item{Try other types of crossover or design a specific crossover for
  our problem.}
\item{Avoid the crossover utilization by using an another optimization
  method, for example the taboo search.} 
\end{itemize}

Finally, this work may give a new way to analyze proteins where the other
methods do not give results.

\vspace{3cm}

\bibliographystyle{IEEEtran}
\bibliography{bibliography}

\begin{thebibliography}{10}
\providecommand{\url}[1]{#1}
\csname url@rmstyle\endcsname
\providecommand{\newblock}{\relax}
\providecommand{\bibinfo}[2]{#2}
\providecommand\BIBentrySTDinterwordspacing{\spaceskip=0pt\relax}
\providecommand\BIBentryALTinterwordstretchfactor{4}
\providecommand\BIBentryALTinterwordspacing{\spaceskip=\fontdimen2\font plus
\BIBentryALTinterwordstretchfactor\fontdimen3\font minus
  \fontdimen4\font\relax}
\providecommand\BIBforeignlanguage[2]{{%
\expandafter\ifx\csname l@#1\endcsname\relax
\typeout{** WARNING: IEEEtran.bst: No hyphenation pattern has been}%
\typeout{** loaded for the language `#1'. Using the pattern for}%
\typeout{** the default language instead.}%
\else
\language=\csname l@#1\endcsname
\fi
#2}}

\bibitem{APP1999}
D.~Perkins, D.~Pappin, D.~Creasy, and J.~Cottrel, ``Probability-based protein
  identification by searching sequence databases using mass spectrometry
  data,'' \emph{Electrophoresis}, vol.~20, pp. 3551--3567, 1999.

\bibitem{AGMG1999}
R.~Gras, M.~M\"uller, E.~Gasteiger, P.~B. S.~Gay, W.~Bienvenut, C.~Hoogland,
  J.~Sanchez, A.~Bairoch, D.~Hochstrasser, and R.~Appel, ``Improving protein
  identification from peptide mass fingerprinting through a parameterized
  multi-level scoring algorithm and an optimized peak detection,''
  \emph{Electrophoresis}, vol.~20, pp. 3535--3550, 1999.

\bibitem{AMB2005}
F.~Monigatti and P.~Berndt, ``Algorithm for accurate similarity measurements of
  peptide mass fingerprints and its application,'' \emph{American Society for
  Mass Spectrometry}, vol.~16, pp. 13--21, 2005.

\bibitem{ABE2001}
V.~Bafna and N.~Edwards, ``Scope: A probabilistic model for scoring tandem mass
  spectra against a peptide database,'' \emph{Bioinformatics}, vol.~1, no.~1,
  pp. 1--9, 2001.

\bibitem{AMMC2004}
J.~Magnin, A.~Masselot, C.~Menzel, and J.~Colinge, ``{OLAV}-{PMF}: a novel
  scoring scheme for high-throughput peptide mass fingerprinting,''
  \emph{Journal of Proteome Research}, vol.~3, pp. 55--60, 2004.

\bibitem{ADAC1999}
V.~Dancik, T.~Addon, and K.~Clauser, ``De novo peptide sequencing via tandem
  mass spectrometry,'' \emph{Journal of Computational Biology}, vol.~6, no.
  3/4, pp. 327--342, 1999.

\bibitem{AFP2005}
A.~Frank and P.~Pevzner, ``Pepnovo: de novo peptide sequencing via
  probabilistic network,'' \emph{Analytical Chemistry}, vol.~77, pp. 964--973,
  2005.

\bibitem{AMHBJC}
J.~Marlard, A.~Heredia-langner, D.~B. K.H., Jarman, and W.~Cannon,
  ``Constrained de novo peptide identification via multi-objective
  optimization,'' in \emph{International Parallel and Distributed Processing
  Symposium}, 2004, p. 191a.

\bibitem{ASD2004}
B.~Searle, S.~Dasari, M.~Turner, A.~Reddy, D.~Choi, P.~Wilmarth, A.~McCormack,
  L.~David, and S.~Nagalla, ``High-throughput identification of proteins and
  unanticipated sequence modifications using a mass-based alignment algorithm
  for {MS}/{MS} de novo sequencing results,'' \emph{Analytical Chemistry},
  vol.~76, pp. 2220--2230, 2004.

\bibitem{AHLR2004}
A.~Heredia-Langner, W.~Cannon, K.~Jarman, and K.~Jarman, ``Sequence
  optimization as an alternative to de novo analysis of tandem mass
  spectrometry data,'' \emph{Bioinformatics}, vol.~20, no.~14, pp. 2296--2304,
  2004.

\bibitem{hollan:ga}
J.~Holland, \emph{Adaptation in Natural and Artificial Systems}.\hskip 1em plus
  0.5em minus 0.4em\relax University of Michigan Press, 1975.

\bibitem{ARVO1995}
A.~Rockwood, S.~V. Orden, and R.~Smith, ``{R}apid {C}alculation of {I}sotope
  {D}istribtion,'' \emph{Analytical Chemistry}, vol.~67, no.~15, pp.
  2698--2704, 1995.

\bibitem{AHH1992}
S.~Henikoff and J.~Henikoff, ``Amino acid substitution matrices from protein
  blocks,'' \emph{Proceedings of the National Academy of Sciences}, vol.~89,
  pp. 10\,915--10\,919, 1992.

\bibitem{davis:adapt}
L.~Davis, ``Adapting operator probabilities in genetic algorithms.'' in
  \emph{Third International Conference on Genetic Algorithms}, J.~D. Schaffer,
  Ed.\hskip 1em plus 0.5em minus 0.4em\relax Morgan Kaufmann, 1989, pp. 61--69,
  san Mateo, CA.

\bibitem{julsrtom:adapt}
B.~A. Julstrom, ``What have you done for me lately? adapting operator
  probabilities in a steady-state genetic algorithm.'' in \emph{Proceedings of
  the sixth International Conference on Genetic Algorithms}, L.~J. Eshelman,
  Ed.\hskip 1em plus 0.5em minus 0.4em\relax Morgan Kaufmann, 1995, pp. 81--87,
  san Francisco, CA.

\bibitem{herrera96adaptation}
\BIBentryALTinterwordspacing
rancisco Herrera and M.~Lozano, ``Adaptation of genetic algorithm parameters
  based on fuzzy logic controllers,'' in \emph{Genetic Algorithms and Soft
  Computing}, F.~Herrera and J.~L. Verdegay, Eds.\hskip 1em plus 0.5em minus
  0.4em\relax Heidelberg: Physica-Verlag, 1996, pp. 95--125. [Online].
  Available: \url{citeseer.nj.nec.com/104774.html}
\BIBentrySTDinterwordspacing

\bibitem{mut:hong}
T.~P. Hong, H.~Wang, and W.~Chen, ``Simultaneosly applying multiple mutation
  operators in genetic algorithms,'' \emph{Journal of Heuristics}, vol.~6, pp.
  439 -- 455, 2000.

\bibitem{Cahon04a}
S.~Cahon, N.~Melab, and E.-G. Talbi, ``Paradis{EO}: {A F}ramework for the
  {R}eusable {D}esign of {P}arallel and {D}istributed {M}etaheuristics,''
  \emph{Journal of Heuristics}, vol.~10, no.~3, pp. 357--380, May 2004.

\end{thebibliography}
\end{document}